\documentclass[aps,pre,twocolumn,superscriptaddress]{revtex4}

\usepackage{amsmath}
\usepackage{graphicx}
\usepackage{wasysym}

\begin{document}

\title{Quasicrystalline Order in Binary Dipolar Systems}

\author{Falk Scheffler} 
\affiliation{Fachbereich Physik, Universit\"at Konstanz, 78457
  Konstanz, Germany}

\author{Philipp Maass} 
\affiliation{Institut f\"ur Physik, Technische Universit\"at Ilmenau,
  98684 Ilmenau, Germany}
\email{Philipp.Maass@tu-ilmenau.de}

\author{Johannes Roth}
\affiliation{Institut f\"ur Theoretische und Angewandte Physik,
  Universit\"at Stuttgart, 70550 Stuttgart, Germany}

\author{Holger Stark}
\affiliation{Fachbereich Physik, Universit\"at Konstanz, 78457
  Konstanz, Germany}

\date{April 15, 2004}

\begin{abstract}
  Motivated by recent experimental findings, we investigate the
  possible occurrence and characteristics of quasicrystalline order in
  two-dimensional mixtures of point dipoles with two sorts of dipole
  moments. Despite the fact that the dipolar interaction potential
  does not exhibit an intrinsic length scale and cannot be tuned a
  priori to support the formation of quasicrystalline order, we find
  that configurations with long--range quasicrystallinity yield minima
  in the potential energy surface of the many particle system. These
  configurations emanate from an ideal or perturbed ideal decoration
  of a binary tiling by steepest descent relaxation. Ground state
  energy calculations of alternative ordered states and parallel
  tempering Monte-Carlo simulations reveal that the quasicrystalline
  configurations do not correspond to a thermodynamically stable
  state. On the other hand, steepest descent relaxations and
  conventional Monte-Carlo simulations suggest that they are rather
  robust against fluctuations. Local quasicrystalline order in the
  disordered equilibrium states can be strong.
\end{abstract}

\pacs{42.70.Qs,42.25.Bs,42.55.Ah,61.44.Br}

\maketitle

\section{Introduction}

Since the surprising discovery of sharp diffraction images with
non--crystallographic symmetry in some rapidly quenched metal alloys
by Shechtman {\it et al.} in 1984 \cite{Shechtman/etal:1984},
quasicrystalline structures have attracted fastly growing interest as
an alternative type of structure of solid matter. As a distinctive
feature, these quasicrystals possess long-range positional order in
combination with a crystallographically 'forbidden' (e.g.\ fivefold)
point group symmetry, which necessarily means aperiodic order. By now,
many systems with a quasicrystalline order have been identified in
nature, most of them being ternary or, in a few cases, binary alloys
(for a review see \cite{Trebin:2003}). Very recently, quasicrystalline
structures with fundamental building blocks much larger than single
atoms have been found in micellar phases of dendrimers (tree--like
molecules)\ \cite{Zeng/etal:2004}. They represent a new mode of
organization in soft matter and are interesting in connection with
photonic bandgap materials\ \cite{Zoorob2000} and photonic
quasicrystal lasers\ \cite{Notomi2004}.

On the theoretical side, the formation of quasicrystals could be
reproduced in simulations of binary mixtures with hard sphere or
Lennard--Jones potentials \cite{Widom/etal:1987,Lancon/etal:1988}. In
these simulations the quasicrystalline structures were stabilized by
tuning the distances corresponding to the minima of the interaction
potentials to match the specific particle--particle distances. 

Recently, some evidence for local patters with fivefold symmetry was
found in two-dimensional binary mixtures of superparamagnetic
colloidal particles
\cite{Zahn1999,Wen/etal:2000,Koenig:2003,Koenig:2004}.  Hence the
intriguing question arises: Can there be quasicrystalline long-range
order in a binary dipolar system despite the fact that the dipolar
interaction potential does not possess tunable intrinsic length
scales?

Here, we tackle this problem by investigating the occurrence of
quasicrystalline order in binary mixtures of point dipoles with
varying dipole strengths. As a two-dimensional reference structure
with fivefold symmetry, we use the prominent rhombic binary tiling
and decorate it with two types of dipoles. Then we let it relax
mechanically and find that for a certain range of dipolar strength
ratios $D$ the mechanically stable configuration preserves the
long--range quasicrystalline order (see sec.~\ref{subsec:mecheq}
below). The final configuration corresponds to a local minimum in the
potential energy surface of the many-particle system. However, its
global features are still undetermined. It could correspond to the
ground state, a thermodynamically stable state within a certain
parameter regime (of temperature, mixing ratio and dipolar strength
ratio), or a metastable state that, below some freezing temperature,
becomes separated from other states by free energy barriers which are
infinite in the thermodynamic limit of infinite system size. This is
reminiscent to the behavior of mean--field spin--glass models
\cite{Mezard/etal:1987}. In both cases, one would expect the dipolar
system to evolve in time at some finite temperature and ultimately
build up permanent long-range quasicrystalline order if its initial
configuration belongs to the attraction basin of the quasicrystalline
state. On the other hand, the mechanically relaxed quasicrystal
structure could correspond to a metastable state from which the system
escapes when it surmounts a finite free energy barrier. Nevertheless,
within the last scenario, it would be interesting to see, whether
quasicrystalline ordering exists locally in the disordered equilibrium
state of the system.

To evaluate the global features as discussed in the last paragraph, we
perform a number of investigations. We first compare in
sec.~\ref{sec:en-calc} the energy of several plausible alternative
ground state structures with the energy of the quasicrystalline state.
Then, we assess in sec.~\ref{sec:steepest-descent} the stability of
the quasicrystalline structure, for varying dipole strength ratio,
with respect to small random perturbations of the particle positions
in the ideal decoration. Finally, in sec.~\ref{sec:mc-sim} we perform
Monte Carlo simulations to analyze the characteristics of
quasicrystalline order at finite temperatures. Since for the
thermalization of the system conventional Monte Carlo techniques
turned out to be ineffective, we applied a parallel tempering protocol
to reach thermodynamic equilibrium.

\section{Model}

\subsection{System parameters and order parameter}

Corresponding to the experimental situation
\cite{Wen/etal:2000,Koenig:2003}, we consider particles moving in a
plane with their dipolar moments all pointing in the same direction
perpendicular to the plane. In experiments this can be realized by
letting superparamagnetic particles float on a liquid meniscus in an
external magnetic field.

\begin{figure} 
\centering
\includegraphics[width=0.48\textwidth]{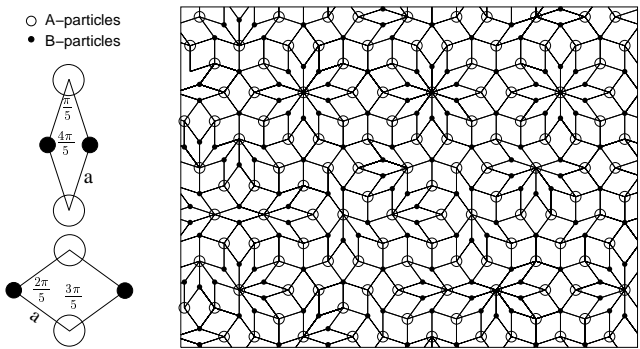}
\caption{The binary tiling decorated by strong (A) and weak (B) dipoles. 
  The angles enclosed by the edges of its two rhombic building blocks
  are all multiples of $\pi/5$, which gives rise to the
  non--translation invariant $10$--fold symmetry.}
\label{fig:structure}
\end{figure}

To investigate quasicrystalline ordering, we choose as a reference
structure the two-dimensional binary tiling, which consists of two
types of rhombs put together by certain matching rules
\cite{Gaehler/etal:1994}. It is a typical example for a
quasicrystalline pattern with fivefold symmetry (see
fig.~\ref{fig:structure}). A crystal results from a periodically
repeated unit cell decorated by atoms. Here we obtain the ideal
quasicrystalline reference structure by decorating the rhombs as
illustrated in fig.~\ref{fig:structure} and described in
ref.~\cite{Lancon/Billard:1986} so that it consists of $N_{\rm A}$
strong A and $N_{\rm B}$ weak B dipoles. The mixing ratio $x=N_{\rm
  A}/(N_{\rm A}+N_{\rm B})$ is $\tau/(2+\tau)\cong0.447$, where
$\tau=(1+\sqrt{5})/2$ is the golden mean. A picture of the final
structure is shown in fig.~\ref{fig:structure}.  The decoration was
already used in previous simulations of binary Lennard--Jones and hard
sphere systems
\cite{Lancon/Billard:1986,Widom/etal:1987,Lancon/etal:1988}. However,
in contrast to these simulations, the potential of the dipolar system
cannot be optimized in an obvious way to support the formation of the
quasicrystalline structure.

The interaction energy of two parallel magnetic moments $m_1$ and $m_2$ at a
distance $r$ is given by
\begin{equation}
E_{\rm int}=\frac{\mu_0}{4\pi}\frac{m_1 m_2}{r^3}\, , 
\label{eq:dipinteract}
\end{equation}
where $\mu_{0}$ is the vacuum permeability. In the following we switch
to dimensionless quantities. As unit length we choose the edge $a$ of
a rhomb, i.e., the distance of two neighboring A-- and B--dipoles in
the reference structure. As unit of energy (and temperature $T$) we
choose the interaction energy $(\mu_0/4\pi)\cdot(m_A m_B /a^3)$ of two
such dipoles. After introducing the ratio $D\equiv m_{\rm A}/m_{\rm
  B}$ of dipole strengths $m_{\rm A}$ and $m_{\rm B}$, the interaction
potentials of two A--dipoles, an A-- and a B--dipole, and two
B--dipoles at distance $r$ are
\begin{equation}
E_{\rm AA}=Dr^{-3}\,,\hspace{1.5em} E_{\rm AB}=r^{-3}\,,\hspace{1.5em}
E_{\rm BB}=D^{-1} r^{-3}\,.
\label{eq:dipinteract-dimless}
\end{equation}
The dimensionless form clarifies that the dipole strength ratio $D$ is
the only additional parameter in our problem besides the temperature
$T$ and the (fixed) mixing ratio $x$. This is a direct consequence of
the scale--free dipolar interaction potential.

In order to quantify the degree of quasicrystalline order, we
define the order parameter
\begin{equation}
\phi=\Bigl| \frac{1}{2N_p}
  \sum_{j,k} \theta(r_{\rm max}-r_{j,k})\,
\exp(i\cdot 10\,\alpha_{j,k})\Bigr|\,,
\label{eq:orderparameter}
\end{equation}
where $\theta(.)$ is the conventional step function ($\theta(x)=1$ for
$x>0$ and zero else), $N_p=\sum_{j,k}\theta(r_{\rm max}-r_{j,k})$ is
the number of pairs of dipoles with distances $r_{j,k}$ smaller than
$r_{\rm max}\equiv1.15$, and $\alpha_{j,k}\equiv\text{\varangle} ({\bf
  r}_{j,k},{\bf\hat{e}})$ is the bond angle between the pair vector
${\bf r}_{j,k}$ and an arbitrary but fixed direction ${\bf\hat{e}}$.
Since the bond--angles in the binary tiling are all multiples of
$\pi/5$ (cf.\ fig.~\ref{fig:structure}), $\phi=1$ in the ideal
quasicrystalline structure, while $\phi=0$ for a system without
tenfold bond orientational order. The value $r_{\rm max}$ is slightly
smaller than the distance of the two A particles in the fat rhomb
shown in fig.~\ref{fig:structure}. Eliminating these pairs from the
sum in eq.~(\ref{eq:orderparameter}), on the one hand, allows us to
encompass slightly displaced ``nearest neighbors'' but, on the other
hand, guarantees that bond angles different from multiples of $\pi/5$
are not counted in the ideal quasicrystalline configuration.

In addition to (\ref{eq:orderparameter}), we also consider the
$n$--fold local bond orientational order parameters
\begin{equation}
\tilde{\phi}_n=\frac{1}{2N} \sum_j \frac{1}{N_j}
\Bigl| \sum_k \theta(r_{\rm max}-r_{j,k})\,
\exp(i\cdot n\,\alpha_{j,k})\Bigr|\,,
\label{eq:alt-orderparameter}
\end{equation}
where $N_j=\sum_k\theta(r_{\rm max}-r_{j,k})$ is the number of
neighbors of dipole $j$. Since the absolute value is taken before
averaging over all $N$ dipoles, the $\tilde{\phi}_n$ are sensitive to
the $n$--fold symmetric arrangement of nearest neighbors around any
dipoles, but insensitive to the spatial variation of the bond
orientations. Naturally, we have $\tilde{\phi}_{10}\geq\phi$.

\subsection{Rational approximants}
\label{subsec:approximants}

Due to the missing translational invariance, standard periodic
boundary conditions cannot be imposed to ideal quasicrystalline
systems. To resolve this problem, it is convenient to use rational
approximants of quasicrystals as described in
\cite{Entin-Wohlman/etal:1988}. A rational approximant is a rectangular
part of the ideal quasicrystalline structure which is chosen such that
it may be exposed to periodic boundary conditions without too much
distorting the local quasicrystalline ordering.

From a number of different rational approximants, we mainly used a
small one with $890$ dipoles ($398$ A, $492$ B dipoles) and size
$24.80 \times 29.15$, and a large one with $1700$ dipoles ($760$ A,
$940$ B dipoles) and size $34.27 \times 40.29$. For both approximants,
we find the order parameter $\phi\simeq0.9998$ close to the ideal
value $\phi=1$. We have no indication that the type of approximant is
decisive for the results obtained below.

To speed up the computer simulations, we store the dipolar interaction
energies (and the forces) in a large matrix by discretizing the set of
possible distances ${\bf r}_{i,j}$ and use a linear interpolation
scheme between the matrix entries. As an advantage, the matrix has to
be calculated only once for all the simulations, and the computer
memory access is usually much faster than the repeated evaluation of
the original mathematical expressions.  Details of the method and how
to take into account the periodic boundary conditions are outlined in
the appendix.

As an example, for the two approximants described above we use a $4044
\times 4754$ matrix for the energies and a $2022 \times 2377$ matrix
for the forces. We find the underlying discretization of space for the
energies or forces on a length scale of order $10^{-2}$--$10^{-3}$ to
have no noticeable influence on our results.

\subsection{Mechanical equilibrium}\label{subsec:mecheq}

When considering the ideal quasicrystalline structure as a potential
(meta--)stable state, the first question one should ask is whether
this structure can be mechanically stable, i.e., whether the net
forces on all the dipoles balance out or, in other words, whether the
structure is a local or global minimum of the systems' potential
energy \cite{saddle-comment}. While in our case for common periodic
lattices, this is obvious from simple symmetry arguments, the
situation is considerably more involved in the quasicrystalline
structure, which is another instance of the peculiarities of this
unusual type of ordering.

Note that in an infinite quasicrystal two different positions are
never exactly equivalent. So the calculation of the long--range
interaction with its surrounding dipoles can, strictly speaking, not
be based on a finite piece of the ideal structure.  Practically,
however, as the dipolar interactions (\ref{eq:dipinteract}) or
(\ref{eq:dipinteract-dimless}) decay as $r^{-3}$ with the distance
$r$, the potential felt the dipoles will largely be dominated by their
local surroundings (see also next section). In this context, it is
interesting to note that in the infinite binary tiling any
arbitrarily large piece of it repeats (up to rotations) within a
distance of the order of its size (see e.g.\ \cite{Baake:1999}).

In the ideal structure, the net forces acting on a dipole converge to
a relative precision of $10^{-6}$ when taking into account neighboring
particles up to a distance of $r_{\max}\simeq 25$.  The resulting
force components clearly show that there is no dipole strength ratio
$D$ which would make the ideal structure a minimum of the potential
energy surface. On the other hand, when we let the ideal
quasicrystalline structure relax via the method of steepest descent
into a local potential minimum, we find that in the range $4\lesssim D
\lesssim 6.5$ the positions of the dipoles are only very slightly
shifted. Based on the steepest descent algorithm, however, we cannot
state simple systematic rules for the decoration of the binary tiling
such that the dipoles assume an exact mechanical equilibrium. Due to
this fact and since the energy per dipole of the relaxed
configurations and the ideal quasicrystal are almost the same, we
retain the decoration of the binary tiling as model reference
structure.

\section{Ground state calculations}
\label{sec:en-calc}

Considering the quasicrystalline structure as a potential equilibrium
state, we next investigate how this structure compares energetically
to plausible alternative ordered states. To this end, we examine the
energy per dipole in a binary dipolar system with the mixing ratio
$x=\tau/(2+\tau)$ and the total number density of dipoles $\rho\simeq
1.231$, which are the values of the ideal quasicrystalline structure.
The dipole strength ratio $D$ then remains as the only free parameter.

To determine the energy per dipole $E_{\star}$ in the ideal
quasicrystalline structure, we calculate the average energy of about
$10^4$ dipoles within a maximum distance $r_{\rm max}=300$ and
extrapolate for $r_{\rm max}\rightarrow \infty$
\cite{extrapol-comment}. In view of the discussion in the previous
section, we should mention that the value of $E_{\star}$ may be up to
a few per mille higher than the energy per dipole in the slightly
'relaxed' quasicrystal. Note that the conclusion from this section
will not alter due to such differences.

The irrational mixing ratio $x$ requires alternative ordered
structures based on regular lattices to be phase--separated, i.e., to
be a combination of two distinct lattices, one with a higher mixing
ratio than in the quasicrystal and another one with a lower mixing
ratio. In the thermodynamic limit, we can neglect energy contributions
from the interfacial boundaries between the two phases and optimize
their respective lattice constants (or their 'volume fractions') to
minimize the total energy per particle $E_i$ of the two-phase state.

\begin{figure} 
\centering
\includegraphics[width=0.48\textwidth]{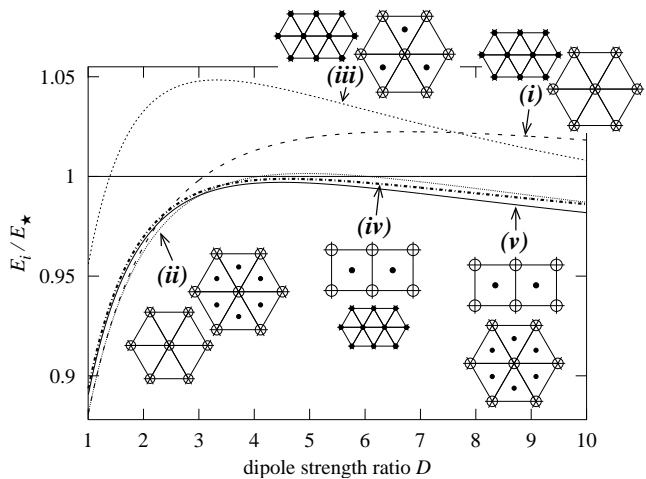}
\caption{Alternative ordered structures and their energies per particle 
  ($E_i$) compared to that of the quasicrystalline structure
  ($E_{\star}$).  Plotted is the fraction $E_i/E_{\star}$ as a
  function of the ratio of dipole moments $D$.}
\label{fig:alt-structures}
\end{figure}

Some plausible alternative structures are depicted in
fig.~\ref{fig:alt-structures}: {\it (i)} two hexagonal lattices (phase
separation of A and B dipoles), {\it (ii)} a hexagonal A lattice with
B dipoles in the triangle centers and a hexagonal A lattice, {\it
  (iii)} a hexagonal A lattice with B dipoles in every second triangle
center and a hexagonal B lattice, {\it (iv)} a centered square lattice
of A and B dipoles and a hexagonal lattice of B dipoles, and {\it(v)}
a centered square lattice of A and B dipoles together with the
centered A--B hexagonal lattice of {\it (ii)}. The corresponding
values $E_i$ for dipole strength ratios $1\leq D\leq 10$ are shown in
the graph in fig.~\ref{fig:alt-structures} relative to the value
$E_{\star}$ of the quasicrystalline structure.

The figure shows the preference of hexagonal order and the degeneracy
of structures {\it (i)} and {\it (ii)} for $D=1$, i.e., identical A
and B dipoles. Furthermore, hexagonal ordering is preferred for
$D\lesssim 3$, whereas for larger $D$, partial tetragonal ordering
seems energetically favorable. The quasicrystalline structure is
closest to the optimum structure in the range $4\lesssim D\lesssim 6$,
with an optimum around $D\simeq 4.5$. However, the phase--separated
structure {\it (v)} clearly has lower energy. Accordingly, the
quasicrystalline structure cannot be the ground state of the dipolar
mixture.

Still, we find that from a purely energetic point of view
the quasicrystalline structure is a surprisingly competitive type of 
ordering  for a binary mixture of dipoles within the appropriate 
mixing ratio and dipole strength ratio.
In view of the fact that the alternative
ordered structures {\it (i)}--{\it (v)} require a phase boundary,
which possesses a surface energy, the quasicrystal may become the
preferred structure in finite systems. Interestingly, the quasicrystal
becomes more favorable when the interaction potential $E_{\rm
  int}\sim r^{-3}$ is modified to a fictitious $\tilde{E}_{\rm
  int}\sim r^{-\alpha}$ with $\alpha$ close to $2$ ($\alpha>2$ for
reasons of convergence). Albeit we are not aware of any realization of
such a potential in nature, this indicates that other types of
scale-free interactions more strongly support the quasicrystalline
ordering.

\section{Stability analysis of reference structures}
\label{sec:steepest-descent}

\subsection{Steepest descent calculations}

In this section, we test the dynamical stability of the
quasicrystalline structure using the method of steepest descent. We
displace the dipoles in small steps along their potential gradient
with the step length being proportional to the modulus of the gradient
vector. In numerics, this algorithm is used to find minima of
multi--parametric functions \cite{num-rec}. In physical terms, it
describes the overdamped motion of the particles at zero temperature,
which may be considered the simplest type of dynamics to be
implemented in a system. The method has been applied before to
Lennard--Jones quasicrystals \cite{Roth/etal:1990}.

In our simulations, the maximum step length (for the dipole
experiencing the largest force) is limited to a fixed value of order
$0.01$--$0.001$. Typically, after $10^3$--$10^4$ steps, the relaxation
is finished when the dipoles start to oscillate about fixed positions,
where the amplitudes are of the order of the maximum step length. We
perform two types of steepest descent calculations for varying dipole
strength ratio $D$. In the first one, we start from the ideal
quasicrystalline positions of the dipoles, and in the second one, we
perturb the ideal positions by Gaussian random noise of different
strength.

The results of these simulations are somewhat hard to quantify since
their concrete numerical outcome depends on the details of the
algorithmic implementation, e.g., the value of the step length. The
long--range dipolar potential and the lack of simple translational
symmetry (which would let forces balance out trivially) give rise to a
large number of dipole configurations being -- mostly shallow -- local
energetic minima. Which of the different minima will be reached in a
steepest descent calculation depends on computational details and
initial conditions. Nonetheless, robust trends in the relaxation
behavior are revealed and we obtain a reliable picture of the systems'
dynamical stability.

\subsection{Stability as a function of dipole strength ratio $D$}

\begin{figure} 
\centering
\includegraphics[width=0.48\textwidth]{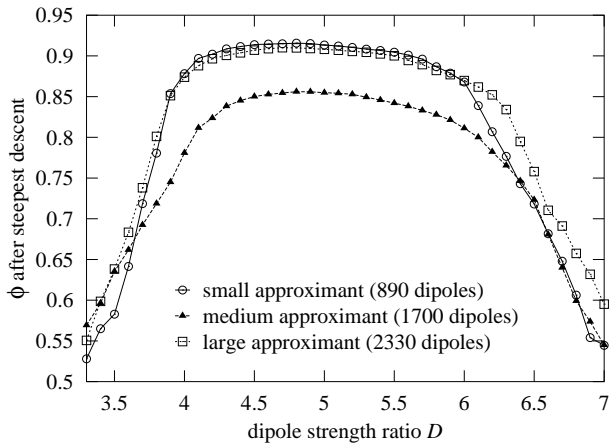}
\caption{Order parameter $\phi$ after steepest descent calculations 
  starting from the ideal quasicrystalline structure as a function of
  the dipole strength ratio $D$ for three different approximants.}
\label{fig:sd-d}
\end{figure}

Figure~\ref{fig:sd-d} shows the order parameter $\phi$ as a function
of the dipole strength ratio $D$ as obtained after steepest descent
calculations starting from the ideal quasicrystalline structure. The
data indicate that the 'relaxed' configurations stay closest to the
ideal structure in a range $4.5\leq D\leq 5.5$ with an optimum around
$D\simeq 4.8$.  The optimum range of $D$ essentially coincides with
the range deduced from the comparison with alternative ordered
structures in sec.~\ref{sec:en-calc}.

This result is reproducible for different values of the step length
and for different approximants (see fig.~\ref{fig:sd-d}), though the
values of the relaxed order parameters $\phi$ may differ.  This
difference is not particularly significant, since the magnitude of
$\phi$ is very sensitive even to small displacements of the dipoles.
For example, we find $\phi\simeq 0.8$ if the dipole coordinates are
randomly altered with respect to their values in the ideal
quasicrystalline structure by a Gaussian noise with standard deviation
$\sigma=0.02$ (cf.\ fig.~\ref{fig:sd-gauss}). Therefore, it is
reasonable to consider the 'relaxed' structures to represent the ideal
quasicrystalline configuration.

\subsection{Recovery from perturbations}

\begin{figure} 
\centering
\includegraphics[width=0.48\textwidth]{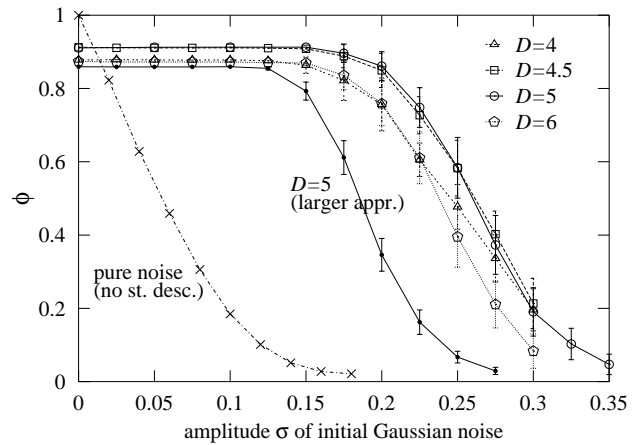}
\caption{Average order parameter $\phi$ after steepest descent calculations 
  starting from a perturbed structure for different values of $D$.
  $\phi$ is plotted as a function of the standard deviation $\sigma$
  of the initial Gaussian displacements. As indicated, one curve
  corresponds to a larger approximant ($1700$ instead of $890$
  dipoles). For comparison, the order parameter $\phi$ due to pure
  Gaussian noise is shown, i.e., before the steepest descent
  relaxation starts.}
\label{fig:sd-gauss}
\end{figure}

Next, we test for metastability of the structure and ask if the
quasicrystalline order recovers from small perturbations. Therefore,
we apply Gaussian noise with zero mean and standard deviation $\sigma$
to each particle coordinate before letting them relax via steepest
descent. The average order parameter $\phi$ of the final
configurations is plotted in fig. \ref{fig:sd-gauss} as a function of
$\sigma$ for a few values of $D$ from the optimum range.

Up to an average initial displacement $\sigma\lesssim 0.15$, the
system relaxes back to the quasicrystalline structure, whereas for
larger $\sigma$, numerous defects remain as reflected by the
decreasing average order parameter $\phi$. This behavior is
reminiscent of the empirical Lindemann criterion, which predicts the
melting of a crystal for an average displacement of $0.14$ in units of
the lattice constant.  We note also that due to the initial Gaussian
noise the order parameter decreases below 0.1 at $\sigma=0.15$ from
which it recovers its optimum value around 0.9. This means that the
quasicrystalline state is relatively robust against random
perturbations.

Figures~\ref{fig:sd-d} and \ref{fig:sd-gauss} also include results for
larger approximants.  Clearly, the increasing system size does not
lead to a significantly enhanced or reduced stability of the
quasicrystalline order.  For example, in Fig.\ \ref{fig:sd-d} the
order parameter at $D=4.8$ is reduced to 0.85 in the medium
approximant but recovers a value above 0.9 in the large approximant.
Hence we conclude that the stability of the quasicrystalline binary
tiling has to be an intrinsic property of the ordered structure rather
than a finite size effect.

\section{Behavior at finite temperatures}
\label{sec:mc-sim}

\subsection{Monte Carlo simulations}

To assess the behavior of the binary system at finite temperatures we
choose Monte Carlo (MC) simulations. While computationally less costly
than e.g.\ Langevin dynamics or even more detailed schemes, the
dynamical MC simulations allow us to explore thermodynamical
equilibrium states and, with a reasonable choice of jump trials,
should also yield a realistic scenario of the systems' evolution.

As a jump trial in our simulations, a particle is chosen at random and
imposed a Gaussian distributed displacement. The trial is accepted
according to the Metropolis rule. The standard deviation of the
Gaussian in the range $0.001$--$0.1$ is adjusted dynamically to ensure
an efficient acceptance rate of trials in the range $10$--$60\%$.

For sufficiently low temperatures, the outcome of these standard MC
simulations goes well together with the results from the steepest
descent simulations. For example, starting from the ideal or slightly
perturbed quasicrystalline structure at $T=0.005$, we find the order
parameter first to decay and then to fluctuate around $\phi\simeq
0.75$, which confirms that (nearly) quasicrystalline order represents
a local energetic minimum in phase space.

\begin{figure} 
\centering
\includegraphics[width=0.48\textwidth]{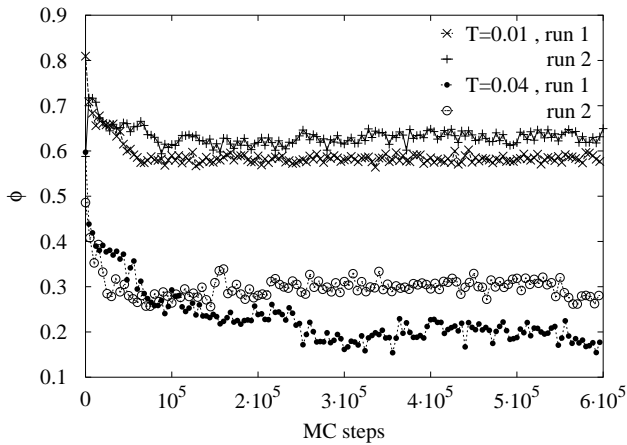}
\caption{Example trajectories of the order parameter $\phi$ as a function 
  time (MC steps) from standard MC simulations at two different
  (fixed) temperatures $T$. The simulations are for $D=5$ in an
  approximant of $760$ dipoles. For each $T$, 'run $1$' started from the
  ideal quasicrystalline structure, while the initial configuration of
  'run $2$' was perturbed by Gaussian noise with $\sigma=0.1$ (cf.
  sec.~\ref{sec:steepest-descent}).}
\label{fig:mc-op-t}
\end{figure}

However, especially at slightly higher $T$, the degree and speed of
the order parameter relaxation vary strongly from run to run in the MC
simulations. In fig.~\ref{fig:mc-op-t}, example trajectories of $\phi$
are shown for $T=0.01$ and $0.04$.  The quasicrystalline structure is
relatively stable at $T=0.01$, but the attained value of $\phi$
depends erratically on the initial conditions of the simulation.  For
example, in fig.\ \ref{fig:mc-op-t} the order parameter emerging from
an initially perturbed structure (run 2) unexpectedly exceeds the
order parameter emerging from the ideal quasicrystalline structure
(run 1). Moreover, at $T=0.04$ a slow and unpredictable decay of
$\phi$ becomes observable on the time scale reached in the simulation.
Even after excessively long runs of more than $10^6$ MC steps, it
remains unclear whether the system has reached an equilibrated state.

\subsection{Parallel tempering}

To accelerate thermalization in the simulations, there are different
possibilities. One of them is to introduce 'artificial' multi particle
flips as additional MC moves, as was previously done for the
Lennard--Jones system in \cite{Widom/etal:1987}. In an extra series of
our simulations, we tested an elementary version of such flips,
whereby every $100$ MC steps the positions of an A and a B dipole are
interchanged and the new configuration is evolved for several MC steps
before the whole move is either accepted or rejected.  In general, we
find the relaxation of $\phi$ to be faster, but the overall behavior
remains unaltered.

For the main part of our simulations, we use the standard local moves
of single particles, which might be closer to the dynamics of the
experimental systems, but we apply a more sophisticated thermalization
scheme known as parallel tempering
\cite{Hukushima/etal:1996,Newman/Barkema:1999}. Basically, the idea of
the method is to circumvent trapping of a systems' dynamics in local
energetic minima at low temperatures by occasionally interchanging the
configuration with the one of the same system simulated in parallel at
higher temperatures. Here, we consider $26$ copies of our system at
temperatures $2.5 \times 10^{-4}=T_1>T_2>\ldots>T_{26}=0.071$. Every
$2000$ MC steps, the particle configurations at adjacent temperatures
$T_i$, $T_{i+1}$ are interchanged with a Metropolis type rate,
\begin{equation}
w_{i,i+1} \!=\!
\begin{cases}
  \exp\left[ \left( \frac{1}{T_i}-\frac{1}{T_{i+1}} \right)\! 
(E_{i+1}-E_i) \right] &\text{if }  E_{i+1}>E_i \\
  \quad\quad\quad\quad\quad\quad 1 &\text{else}\,.
\end{cases}
\label{eq:pt-metro}
\end{equation}
where $E_i$ and $E_{i+1}$ are the energies of the configurations. It
can be shown that this scheme allows different configurations to occur
with their correct Boltzmann weight at any of the temperatures $T_i$.

In our implementation of the algorithm, a control program on a single
PC keeps track of the configurations simulated at the various
temperatures $T_i$. Once the assignment of a certain configuration to
a $T_i$ has been made for the next $2000$ MC steps, it is passed as
independent computing job to our queuing system. So the simulations
can be carried out on a variable number of available CPUs which may
also differ in speed. One MC step for a single configuration takes
about $0.5$\,s on a contemporary Intel Pentium IV $2.8$\,GHz CPU.

We find the parallel tempering algorithm to be highly effective in
thermalizing our ensemble of $26$ systems. This relatively large
number allows us to keep the spacing between the $T_i$ small and thus
to change configurations frequently while still covering a large range
of temperatures from possible ordering to apparently fluid--like
behavior. After typically $300$ rounds (i.e.\ $300\times2000$ MC
steps), we see no qualitative differences any more between the
extremes of an ensemble started from the ideal quasicrystalline
structure and one started from random initial positions.

\subsection{Results: local ordering}

\begin{figure} 
\centering
\includegraphics[width=0.48\textwidth]{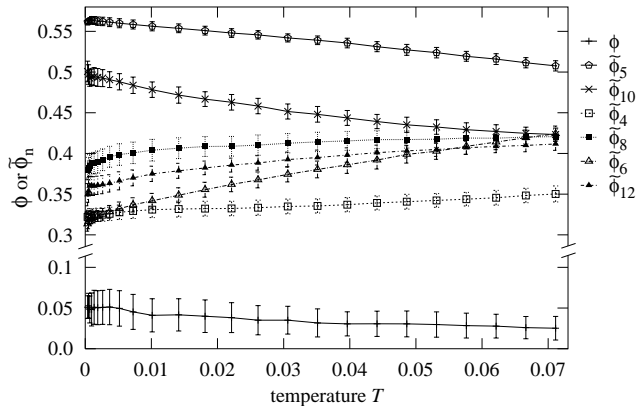}
\caption{Order parameters $\phi$ and $\tilde{\phi}_n$ for several $n$ 
  as a function of temperature $T$ as obtained from the parallel
  tempering MC simulations. The dipole strength ratio is $D=4.75$.
  Note the jump in the ordinate.}
\label{fig:op-vs-T}
\end{figure}

Figure~\ref{fig:op-vs-T} shows the mean order parameters $\phi$ and
$\tilde{\phi}_n$ for several $n$ as a function of temperature $T$. The
values are obtained from particle configurations at the respective
$T_i$, each taken at the end of the $2000$ MC step cycles of the
parallel tempering scheme. The simulations are carried out at a dipole
strength ratio $D=4.75$ and started from random initial positions. The
behavior is very similar for $D=4.5$ and $5$. Apart from larger
initial fluctuations of the order parameters $\phi$ and
$\tilde{\phi}_{10}$, the same holds for simulations started from
ordered initial positions.

\begin{figure} 
\centering
\includegraphics[width=0.415\textwidth]{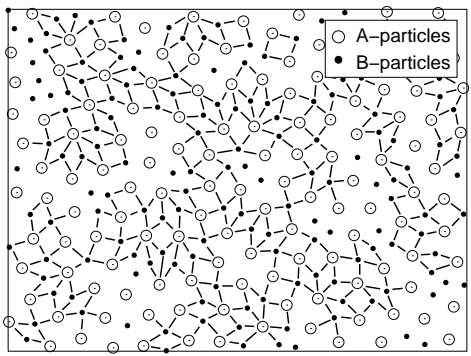}
\caption{Configuration from the parallel tempering MC
  simulations for $D=4.75$. The temperature is $T=0.00134$, the order
  parameter $\phi\simeq 0.07$ and energy per particle $E\simeq 8.087$.
  The part of the system shown is at the same scale as in
  fig.~\ref{fig:structure}. The lines highlight local quasicrystalline
  ordering and are drawn whenever the bond angles between nearest
  neighbors of A (B) particles are even (odd) multiples of $2\pi/10$
  (within an error of 10\%).}
\label{fig:example-structure}
\end{figure}

In fig.\ \ref{fig:example-structure}, part of an example configuration
from the lower temperature range in the parallel-tempering ensemble is
shown.  In this representative example, no long-range quasicrystalline
ordering is discernable. This finding is further supported by
fig.~\ref{fig:op-vs-T}. The order parameter $\phi$ stays very small
throughout the whole temperature range and there is no signature of a
phase transition.

We find that the mean potential energy per particle in the final
configurations approximately fulfills $E(T)\simeq 8.083+T$ at low $T$,
where the simple dependence on $T$ can be understood from a harmonic
approximation around the configurations with lowest energy.  Thus, at
low temperatures, typical configurational energies are below both the
energy of the ideal quasicrystalline structure ($E_{\star}\simeq
8.103$ for $D=4.75$) and the one reached in the steepest descent
calculations, cf.\ sec.~\ref{sec:steepest-descent}. On the other hand,
$E$ is still above the energy of the optimum phase separated structure
($E_{(v)}\simeq 8.078$) found in the ground state energy calculations
in sec.~\ref{sec:en-calc}. This is not surprising, since the necessity
of a phase boundary might prevent the dipolar mixture from reaching
its possible ground state as deduced for infinite system size.

Based on these results, we can exclude the spontaneous occurrence of
long--range quasicrystalline order in a thermodynamically stable
phase. Moreover, irrespective of the (nearly) quasicrystalline
structure being stable against small mechanical perturbations, we
think that it may not represent a thermodynamically metastable state.
Such a state would become separated from other states by an infinite
free energy barrier in the limit of infinite system size.

In accordance with experimental observations
\cite{Wen/etal:2000,Koenig:2003}, we find, however, that, while the
overall structure is amorphous, there occur small domains with
particles arranged as in the quasicrystal, see
fig.~\ref{fig:example-structure}. As can be seen from the value of
$\tilde{\phi}_5$ and $\tilde{\phi}_{10}$ compared to
$\tilde{\phi}_{n}$ with $n=4,8$ and $6,12$ in fig.~\ref{fig:op-vs-T},
there is a tendency towards a preferred local $5$-- or $10$--fold
symmetry at lower temperatures. 

\section{Conclusions}

Our results from the parallel tempering method provide ample evidence
that the quasicrystalline binary tiling with two sorts of dipoles does
not correspond to a thermodynamical equilibrium state.  Nevertheless,
we can identify a range for the dipole strength ratio where a
long--range quasicrystalline structure corresponds to a local minimum
in the potential energy landscape of the system. This local minimum
has an attraction basin covering Gaussian fluctuations of the particle
positions up to 15\% of the distance of two neighboring A-- and
B--dipoles in the ideal reference structure. Our simulations, however,
do not indicate that the barriers separating the local minimum from
other minima increase with the system size. Therefore we do not expect
that such a frozen state with quasicrystalline order exists although
we cannot strictly exclude this possibility \cite{metastab-comment}.

In any case, at low temperatures the times for the system to escape
the local minimum by surmounting free energy barriers can become
rather long. This is clearly seen in the kinetics modeled by
conventional Monte Carlo simulations. Hence structures with
long--range quasicrystalline order can be kinetically stable over
sufficiently long time to make them an interesting subject for further
study and eventual applications. What ``sufficiently long'' in this
context means could be tested in experiments. Todays optical tweezer
techniques allow colloidal particles to be placed at defined
positions. Accordingly, one could prepare a quasicystalline pattern
and monitor its stability.  Moreover, special boundary conditions and
external stimuli may support the formation of quasicrystalline
structures. We also found that a modified scale--free interaction
potential $\propto r^{-(2+\varepsilon)}$ leads, for $\varepsilon>0$
becoming small, to ground state energies of the quasicrystal that
within numerical error bars cannot be distinguished from the most
favorable phase--separated lattice structures investigated in
section~\ref{sec:en-calc}.

In agreement with experiments we find that even in the disordered
ground state of the binary dipole system, local bond orientational
order with quasicrystalline $5$-- and $10$--fold symmetry is preferred. We
made quantitative predictions for the temperature behavior of several
local bond-orientation order parameters, which can be tested in
experiments by using, for example, two-dimensional binary mixtures of
superparamagnetic colloidal particles\ 
\cite{Zahn1999,Koenig:2003,Koenig:2004}.

\begin{acknowledgments}
We gratefully acknowledge fruitful discussions with H.~K\"onig and
G.~Maret and thank the Sonderforschungsbereich 513 of the Deutsche
Forschungsgemeinschaft for financial support.
\end{acknowledgments}

\appendix*

\section{Energy and force calculations in the simulations}

\subsection{General procedure}

For an efficient calculation of the energy $E(x,y)$ of (or force ${\bf
  F}(x,y)=-\boldsymbol{\nabla} E(x,y)$ acting on) a pair of dipoles
$i,j$ with distance vector ${\bf r}_{i,j}=(x,y)$, including their
images in the periodically continued systems of the simulation box, we
store the corresponding values on a fine grid of pair vectors and use
a linear interpolation to obtain the values for ${\bf r}_{i,j}$ in the
continuum. According to the ``minimum image convention'', possible
distances in $x$-- and $y$-- direction fall in the range $-L_x/2 \leq
x< L_x/2$ and $-L_y/2 \leq y<L_y/2$, where $L_x$ and $L_y$ are the
lengths of the system in the $x$-- and $y$--direction. We define by
$\gamma\equiv L_x/L_y$ the aspect ratio. Due to symmetry, $E(x,y)$ is
an even function of $x$ and $y$, while the force components
$F_x(x,y)=-\partial_x E(x,y)$, $F_y(x,y)=-\partial_y E(x,y)$ are odd
functions of $x$, $y$, and even functions of $y$, $x$, respectively.
These symmetries are used to reduce the storage needs for the matrices
of energy values and force components on the grid.

For short notation, we use in this appendix $L_y=1$ as our length unit
and $\mu_0 m_im_j/4\pi L_y^3$ as our energy unit, where $m_i$ and
$m_j$ are the magnetic moments of the two dipoles with pair vector
${\bf r}_{i,j}$ (transformation to the units used in the main text
follows after elementary rescaling). Then we have
\begin{equation}
E(x,y)=\sum_{\mu,\nu=-\infty}^{\infty} \frac{1}{\big[(x+\gamma\mu)^2 + 
  (y+\nu)^2\big]^{3/2}}\, .
\label{eq:e-series}
\end{equation}
The numerical calculation of this absolutely convergent series can be
done by different means, for example by employing a two--dimensional
variant of the Ewald summation \cite{Allen/Tildesley:1990} or a simple
extrapolation scheme, cf.\ \cite{extrapol-comment}. We applied a
method developed previously in our group \cite{Rinn:2000}, where the
series in (\ref{eq:e-series}) is decomposed into an inner part for
distances $\sqrt{(x+\gamma\mu)^2 + (y+\nu)^2}$ smaller than a cutoff
radius $r_{\rm m}$, and a remaining outer part, $E=E_{\rm in}+E_{\rm out}$.
The inner part $E_{\rm in}$ is calculated by explicitly performing the
summation, while the outer part $E_{\rm out}$ is approximated by an
integral. An analogous decomposition is done for the force components.
In the following we discuss the integral approximation for the outer
parts and their numerical evaluation.

\subsection{Integral approximation for the energy}

Defining for $\alpha=0,1$
\begin{equation}
\nu_\alpha(\mu) = 
\begin{cases} 
\operatorname{ceil}\sqrt{r_{\rm m}^2 -\gamma^2\mu^2}
   &\text{if } \gamma|\mu| < r_{\rm m} \\
\quad \alpha & \text{else} \, ,
\end{cases} \\
\end{equation}
where $\operatorname{ceil}(x)$ is the lowest integer number larger
than or equal to $x$, we obtain
\begin{widetext}
\begin{eqnarray}
E_{\rm out} &=& \sum_{\mu=-\infty}^\infty \left[ 
\sum_{\nu=-\infty}^{-\nu_0(\mu)} \left[ \dots \right]^{-3/2}
  +\sum_{\nu=\nu_1(\mu)}^\infty \left[ \dots \right]^{-3/2} \right] \\
&\simeq& \int_{-\infty}^\infty d\mu\, \left[ \int_{-\infty}^{-\nu_0(\mu)}
    d\nu\, \left[ \dots \right]^{-3/2} +
    \int_{\nu_1(\mu)}^\infty d\nu\, 
\left[ \dots \right]^{-3/2} \right. +\nonumber \\ 
&& 
+ \frac{1}{2} \left( \left[ (x+\gamma\mu)^2 + (y -
    \nu_0(\mu))^2 \right]^{-3/2} + \left[ (x+\gamma\mu)^2 + (y +
    \nu_1(\mu))^2 \right]^{-3/2} \right) \Biggr]\\
&\equiv& \int_{-\infty}^\infty d\mu\, f(\mu; x,y)\, .
\label{eq:e-out-int}
\end{eqnarray}

Here $[\dots]$ stands for $\big[(x+\gamma\mu)^2 + (y+\nu)^2\big]$. The two 
integrals over $\nu$ together yield
\begin{equation} 
\int d\nu\, \ldots = \frac{1}{(x+\gamma\mu)^2} \left(
      2 - \frac{\nu_0(\mu) - y}{\sqrt{(x+\gamma\mu)^2 + (y-\nu_0(\mu))^2}}
 -\frac{\nu_1(\mu) + y}{\sqrt{(x+\gamma\mu)^2 + (y+\nu_1(\mu))^2}}
\right)\, .
\end{equation} 

Due to the piecewise definition of $\nu_{\alpha}(\mu)$, it is useful
to split the remaining integral over $\mu$ in (\ref{eq:e-out-int})
according to its boundaries, with
\begin{equation}
\int_{-\infty}^\infty d\mu\, f(\mu;x,y)= 
\int_{|\mu|\leq r_{\rm m}/ \gamma} d\mu\, f(\mu;x,y)
+\int_{|\mu|\geq r_{\rm m}/ \gamma} d\mu\, f(\mu;x,y)
\equiv E_{{\rm out},1}+E_{{\rm out},2} 
\label{eq:e-out-1-2}\, .
\end{equation}
The first part $E_{{\rm out},1}$ is calculated analytically.
With the abbreviations
\begin{equation}
  a_{\pm}\equiv\sqrt{1+\left(\frac{y}{r_{\rm m}\pm x}\right)^2}\,,\hspace{3em}
  b {\pm}\equiv\sqrt{1+\left(\frac{y+1}{r_{\rm m}\pm x}\right)^2},
\end{equation}
it reads
\begin{equation}\hspace{-0.4em}
E_{{\rm out},1} = \frac{1}{\gamma}\left[ \frac{2\, r_{\rm m}}{r_{\rm m}^2-x^2}  
- \frac{2- a_- -a_+}{y} + \frac{2 - b_- - b_+}{y+1} + 
\frac{2 - \frac{1}{a_-} -
   \frac{1}{a_+}}{2 y^2} + \frac{2 - \frac{1}{b_-} - 
\frac{1}{b_+}}{2 (y+1)^2}\right]\,.
\end{equation}
If $y=0$, the limit $y\rightarrow 0$ should be taken
explicitly to avoid numerical instabilities,
\begin{eqnarray}
\lim_{y\rightarrow 0} E_{{\rm out},1} &=& 
\frac{1}{\gamma}\left[ 2 \left( \frac{1}{r_{\rm m}-x} 
+ \frac{1}{r_{\rm m}+x} \right) + (2 - b_- - b_+) +\right. \nonumber \\
&& \phantom{\frac{1}{\gamma}\left[ \right.} 
\left. + \frac{1}{4} \left( \frac{1}{(r_{\rm m}-x)^2} 
+ \frac{1}{(r_{\rm m}+x)^2} \right)
 + \left(1 - \frac{1}{2b_-} - \frac{1}{2b_+} \right)
\right]\,.
\end{eqnarray}

The second part $E_{{\rm out},2}$ is approximated by a sum,
\begin{equation}
E_{{\rm out},2}\simeq \sum_{\mu=-\mu_1+1}^{\mu_1-1} f(\mu;x,y) 
+\left(\frac{1}{2}-\varepsilon\right) \left[ f(-\mu_1;x,y)+f(\mu_1;x,y) 
\right]\, , 
\label{eq:e-out-2-sum}
\end{equation}
where $\mu_1\equiv\operatorname{ceil}(r_{\rm m}/ \gamma)$ and $\varepsilon
\equiv \mu_1-r_{\rm m}/ \gamma$. For numerical stability, here the limit
$x\rightarrow 0$ of the addend with $\mu=0$ should be considered
explicitly,
\begin{equation}
\lim_{x\rightarrow 0} f(0;x,y) = \frac{1}{2}\left[ (r_{\rm m}-y)^{-3}
+ (r_{\rm m}+y)^{-3} + (r_{\rm m}-y)^{-2} + (r_{\rm m}+y)^{-2}\right]\, . 
\end{equation}
\end{widetext}

In the numerics, the evaluation of (\ref{eq:e-out-2-sum}) can
conveniently be combined with the summation for the inner part $E_{\rm
  in}$. For the energy matrix (and for the force calculation discussed
in the next section), we use $r_{\rm m}=10$. If $E$ is to be
calculated repeatedly in a simulation, $r_{\rm m}\simeq 5$ might be a
reasonable choice. In ample tests in the range $0.8\leq\gamma\leq1.2$,
we find the relative error of the integral approximation of the energy
not to exceed $8\times 10^{-4}$ for $r_{\rm m} = 5$ and $10^{-4}$ for
$r_{\rm m} = 10$.

\subsection{Integral approximation for the force}

The outer part $F_x^{\rm out}$ of the $x$ component force is
\begin{widetext}
\begin{eqnarray}
F_x^{\rm out} &=& \int_{-\infty}^{\infty} d\mu\, 
\left[ \frac{1}{x+\gamma\mu} \left(
\frac{y-\nu_{0}(\mu)}{\left[ (x+\gamma\mu)^{2} + (y-\nu_{0}(\mu))^{2}
 \right]^{3/2}} - \frac{y+\nu_{1}(\mu)}{\left[ (x+\gamma\mu)^{2} +
 (y+\nu_{1}(\mu))^{2} \right]^{3/2}} \right) + \right. \nonumber \\
  && {} + \frac{2}{(x+\gamma\mu)^{3}} \left( \frac{y-\nu_{0}(\mu)}{\left[
      (x+\gamma\mu)^{2} + (y-\nu_{0}(\mu))^{2} \right]^{1/2}} -
      \frac{y+\nu_{1}(\mu)}{\left[ (x+\gamma\mu)^{2} + (y+\nu_{1}(\mu))^{2}
      \right]^{1/2}}\right) + {} \nonumber \\
  && \left. {} + \frac{3(x+\gamma\mu)}{2} \left( \frac{1}{\left[ 
(x+\gamma\mu)^{2} +(y-\nu_{0}(\mu))^{2} \right]^{5/2}} +
      \frac{1}{\left[ (x+\gamma\mu)^{2} + (y+\nu_{1}(\mu))^{2} \right]^{5/2}}
    \right) \right]\\
&\equiv& \int_{-\infty}^\infty d\mu\, g(\mu; x,y)\, .
\end{eqnarray}
Analogous to (\ref{eq:e-out-1-2}), the integral over $\mu$ is split
into two parts, of which the first one yields
\begin{eqnarray}
 F_x^{\rm out,1} &=& \frac{1}{\gamma}
\Biggl\{\frac{1}{y} 
    \left[\frac{1}{|r_{\rm m}-x|}\left(\frac{1}{a_-}-a_-\right)
          -\frac{1}{|r_{\rm m}+x|}\left(\frac{1}{a_+}-a_+\right)\right]
\nonumber\\
&&\hspace{0.3em}{}-\frac{1}{y+1} 
   \left[\frac{1}{|r_{\rm m}-x|}\left(\frac{1}{b_-}-b_-\right)
         -\frac{1}{|r_{\rm m}+x|}\left(\frac{1}{b_+}-b_+\right)\right]
\nonumber\\
&&\hspace{0.3em}
{}-\frac{1}{2|r_{\rm m}-x|^3}\left(\frac{1}{a_-^3}+\frac{1}{b_-^3}\right)
+\frac{1}{2|r_{\rm m}+x|^3}\left(\frac{1}{a_+^3}+\frac{1}{b_+^3}
\right)\nonumber\\
&&\hspace{0.3em}
+2\left(\frac{1}{(r_{\rm m}+x)^2}-\frac{1}{(r_{\rm m}-x)^2}\right) 
\Biggr\}\,, 
\end{eqnarray}
while the second part is approximated by a sum analogous to
(\ref{eq:e-out-2-sum}),
\begin{equation}
F_x^{\rm out,2} \simeq \sum_{\mu=-\mu_1+1}^{\mu_1-1} g(\mu;x,y) 
+\left(\frac{1}{2}-\varepsilon\right) \left[ g(-\mu_1;x,y)+g(\mu_1;x,y) 
\right]\, .
\end{equation} 
\end{widetext}

Since the approximation turns out to be inaccurate in the neighborhood
of $x=0$ and $x=0.5\gamma$, it is advantageous to perform the force
calculation in the ranges $|x|/\gamma \lesssim 5\times 10^{-4}$ and
$0.5 - |x|/\gamma \lesssim 10^{-2}$ by explicit summation.

The corresponding expressions for the force component $F_y^{\rm out}$
can be obtained from the expressions for $F_x^{\rm out}$ by
interchanging $x$ and $y$, replacing $\gamma$ by $1/\gamma$, and
rescaling by $1/\gamma^4$, i.e.\ $F_y^{\rm out}(x,y;\gamma)=
1/\gamma^4 \, F_x^{\rm out}(y/\gamma,x/\gamma;1/\gamma)$.


\begin{thebibliography}{9}

\bibitem{Shechtman/etal:1984} D.~Shechtman, I.~Blech, D.~Gratias, and
  J.~W.~Cahn, Phys.~Rev.~Lett.\ {\bf 53}, 1951 (1984).
  
\bibitem{Trebin:2003} H.-R.~Trebin (ed.), {\it Quasicrystals --
    Structure and Physical Properties} (Wiley--VCH, Berlin, 2003).

\bibitem{Zeng/etal:2004} X.~Zeng, G.~Ungar, Y.~Liu, V.~Percec,
  A.~E.~Dulcey, and J.~K.~Hobbs, Nature {\bf 428}, 157 (2004).
  
\bibitem{Zoorob2000} M.~E. Zoorob, M.~D.~B. Charlton, G.~J. Parker,
  J.~J. Baumberg, and M.~C.  Netti, Nature\ \textbf{404}, 740 (2000).
  
\bibitem{Notomi2004} M.~Notomi, H.~Suzuki, T.~Tamamura, and
  K.~Edagawa, Phys.~Rev.~Lett.\ \textbf{92}, 123906 (2004).

\bibitem{Widom/etal:1987} M.~Widom, K.~J.~Strandburg, and
  R.~H.~Swendsen, Phys.~Rev.~Lett.\ {\bf 58}, 706 (1987).

\bibitem{Lancon/etal:1988} F.~Lan\c con and L.~Billard,
  J.~Phys.~France {\bf 49}, 249 (1988).  

\bibitem{Zahn1999} K.~Zahn, R.~Lenke, and G.~Maret, Phys.~Rev.~Lett.\ 
  \textbf{82}, 2721 (1999).

\bibitem{Wen/etal:2000} W.~Wen, L.~Zhang, and P.~Sheng,
  Phys.~Rev.~Lett.\ {\bf 85}, 5464 (2000).

\bibitem{Koenig:2003} H.~K\"onig, Ph.D. thesis, Universit\"at Konstanz,
  2003.

\bibitem{Koenig:2004} H.~König, K.~Zahn, and G.~Maret, 
  in {\it AIP Conference Proceedings: Slow Dynamics in Complex
    Systems}, edited by M.~Tokuyama and I.~Oppenheim (February 2004,
  in press).

\bibitem{Mezard/etal:1987} M.~M\'ezard, G.~Parisi, and M.~A.~Virasoro,
  {\it Spin Glass Theory and beyond} (World Scientific, Singapore,
  1987).

\bibitem{Gaehler/etal:1994} F.~G\"ahler, M.~Baake, and M.~Schlottmann,
  Phys.~Rev.~B\ {\bf 50}, 12458 (1994).

\bibitem{Lancon/Billard:1986} F.~Lan\c con and L.~Billard,
  Europhys.~Lett. {\bf 2}, 625 (1986).

\bibitem{Entin-Wohlman/etal:1988} O.~Entin--Wohlman, M.~Kleman, and
  A.~Pavlovitch, J.~Phys.~France {\bf 49}, 587 (1988). 

\bibitem{saddle-comment} Strictly speaking, zero net force can also
  mean a saddle point in the energy landscape.

\bibitem{Baake:1999} M.~Baake, in: {\it Quasicrystals}, edited by
  J.~B.~Suck, M.~Schreiber, and P.~H\"aussler (Springer, Berlin 2001).

\bibitem{extrapol-comment} When one takes into account neighboring
  dipoles up to a maximum distance $r_{\rm max}$, the energy of an
  individual dipole in the ideal structure scales as
  $E\sim\int^{r_{\rm max}}r^{-3}\, r dr\sim 1/r_{\rm max}$ for large
  $r_{\rm max}$. In the range $50\leq r_{\rm max} \leq 300$ used by
  us, the above scaling relation becomes practically exact and may
  conveniently be used to extrapolate for $r_{\rm max}\rightarrow
  \infty$.

\bibitem{num-rec} see e.g.\ W.~H.~Press, S.~A.~Teukolsky,
  W.~T.~Vetterling, and B.~P.Flannery, {\it Numerical Recipes in C}
  (Cambridge University Press, Cambridge 1995).

\bibitem{Roth/etal:1990} J.~Roth, R.~Schilling, and H.--R.~Trebin,
  Phys.~Rev.~B {\bf 41}, 2735 (1990).

\bibitem{Hukushima/etal:1996} K.~Hukushima and K.~Nemoto,
  J.~Phys.~Soc.~Jpn. {\bf 65} 1604 (1996)

\bibitem{Newman/Barkema:1999} M.~E.~J.~Newman, G.~T.~Barkema, {\it
    Monte Carlo Methods in Statistical Physics} (Clarendon Press,
  Oxford 1999).
  
\bibitem{metastab-comment} Our conclusion is based on the variation of
  the system size in steepest descent and parallel tempering MC
  simulations. Strictly speaking, it would require elaborate standard
  MC simulations (for sufficient statistics) in a large enough variety
  of system sizes, which is beyond the scope of the present work.

\bibitem{Allen/Tildesley:1990} see e.g.\ M.~P.Allen and
  D.~J.~Tildesley, {\it Computer Simulations of Liquids} (Clarendon
  Press, Oxford 1990).

\bibitem{Rinn:2000} B.~Rinn (private communication).

\end{thebibliography}
\end{document}